\documentclass[]{aa}
\usepackage{graphicx}
\usepackage{natbib}
\begin{document}
\topmargin=-1.in
\title{Strong first-order phase transition in a rotating neutron star core and the associated energy release}
\author{J. L. Zdunik\inst{1}  \and
M. Bejger\inst{2,1}
 \and
P. Haensel\inst{1}
 \and
E. Gourgoulhon\inst{2}
}
\institute{N. Copernicus Astronomical Center, Polish
           Academy of Sciences, Bartycka 18, PL-00-716 Warszawa, Poland
{\em jlz@camk.edu.pl}} \institute{N. Copernicus Astronomical
Center, Polish Academy of Sciences, Bartycka 18, PL-00-716
Warszawa, Poland \and LUTh, Observatoire de Paris, CNRS, Universit{\'e} Paris
Diderot, 5 Pl. Jules Janssen, 92190 Meudon, France\\
{\tt jlz@camk.edu.pl, bejger@camk.edu.pl, haensel@camk.edu.pl,
Eric.Gourgoulhon@obspm.fr
 }}
\offprints{J.L. Zdunik}
\date{Received xxx Accepted xxx}
\abstract{}{We calculate the energy release associated with a
strong first-order phase transition, from normal phase
N to an ``exotic''  superdense phase S,  in a rotating neutron star.
Such a phase transition N$\longrightarrow$S, accompanied
by a density jump $\rho_{\rm _N}\longrightarrow \rho_{\rm
_S}$,  is characterized
by $\rho_{_{\rm S}}/\rho_{_{\rm N}}> {3\over
2}(1+P_0/\rho_{_{\rm N}}c^2)$, where $P_0$ is the pressure, at
which phase transition occurs.  Configurations
with small S-phase cores are then unstable and
collapse into stars with large S-phase cores. The energy
release is equal to the difference in
mass-energies between the initial (normal) configuration
and the final configuration containing an S-phase core,
total stellar baryon mass and angular momentum being kept
constant. }{The calculations of the energy release are based
on precise numerical 2-D calculations. Polytropic equations of state
(EOSs) as well as realistic EOS with strong first-order phase transition
 due to kaon condensation  are used. For polytropic EOSs,
a large parameter space is studied.}
{For a fixed ``overpressure'',
$\delta\overline{P}$, defined as the relative excess of
central pressure of collapsing metastable star over the
pressure of equilibrium first-order phase transition, the
energy release $E_{\rm rel}$ does not depend on the
stellar angular momentum. It  coincides with that for
nonrotating stars with the same $\delta\overline{P}$. Therefore,
  results of 1-D calculations of $E_{\rm rel}(\delta\overline{P})$
  for non-rotating stars can be used to predict, with very
  high precision, the outcome of much harder to perform  2-D
  calculations for rotating stars with the same
  $\delta\overline{P}$. This result holds also for
  $\delta\overline{P}_{\rm min}<\delta\overline{P}<0$,
  corresponding to phase transitions  with climbing over the energy barrier
  separating metastable N-phase configurations from
  those with an S-phase core. Such phase transitions could be
  realized in the cores of newly born, hot, pulsating neutron stars. }{}


\titlerunning{Neutron-star phase transition induced core-quakes}
\maketitle
%
\section{Introduction}
\label{sect:introd}
 One of the mysteries of neutron stars is  the actual
 structure of their superdense cores. Many  theories of dense matter
 predict there   a phase transition
  into an ``exotic'' (i.e., not observed
in laboratory) state. Theoretical predictions include boson
condensation of pions and kaons, and deconfinement of quarks
(for review see, e.g., \citealt{Glend.book,weber.book,NSB1}).

The first-order phase transitions, accompanied by discontinuities in the
thermodynamic potentials,  seem to be the
most interesting, as far as the the structure and dynamics of neutron stars
are concerned. In the simplest case, one
considers states consisting of one pure phase. High degeneracy
of the matter constituents implies that the  effects of
temperature can be neglected. In the thermodynamic equilibrium,
the phase transition occurs then at a well defined pressure $P_0$.
It is accompanied by a density jump at the phase interface.

A first-order phase transition in neutron star core is
associated with a collapse of initial  metastable
configurations built of
exclusively of non-exotic (normal - N) phase,   into a more
compact configuration with a core of the superdense (S) exotic
phase. At the core edge the pressure is $P_0$, and the density
undergoes a jump from $\rho_{_{\rm N}}$ on the N-side to
$\rho_{_{\rm S}}$ on the S-side. The collapse, called
``corequake'', is associated with a release of energy,
$E_{\rm rel}$. A neutron star corequake implied by a first-order phase
transition in stellar core could occur during an evolutionary
process,  in which central pressure increases. Examples
of such processes are mass accretion and pulsar spin-down.
In both cases, initial and final
configurations are rotating. One assumes that the baryon mass
of collapsing star, $M_{\rm b}$, is conserved (no mass
ejection) and that the total angular momentum, $J$, is also
conserved ($J$ loss due to radiation during a corequake
due to radiation of the electromagnetic and gravitational waves
is negligible).

Crucial for the corequake is the value of the parameter
$\lambda\equiv \rho_{_{\rm S}}/\rho_{_{\rm N}}$. If
$\lambda<\lambda_{\rm crit}\equiv {3\over 2}
(1+P_0/\rho_{_{\rm N}}c^2)$, then the configurations with
arbitrarily small  cores of the S-phase are stable with respect to
axisymmetric perturbations \citep{Seidov1971,Kaempfer1981,HZS,ZHS}. 
To get a
corequake in an evolving neutron star, a metastable core of
the N-phase, with central pressure $P_{\rm c}>P_0$ and radius
$r_{_{\rm N}}$ should form first. This core is
``overcompressed'', with degree of overcompression measured by
a dimensionless ``overpressure'' $\delta\overline{P}\equiv (P_{\rm
c}-P_0)/P_0$. At some critical value of the overpressure, the
S-phase nucleates, and the S-core of radius $r_{_{\rm S}}$
forms. For $\delta\overline{P}\longrightarrow 0$, we
have $r_{_{\rm
S}}\longrightarrow 0$. This is the case of a {\it weak first
order phase transition}. Up to now, all but one numerical calculations
were restricted to the spherical non-rotating neutron stars
\citep{HaenProsz82,HZS,ZHS,HDPopov1990,MT1992}. 
An exception is a very recent paper,  based on precise
2-D calculations performed for weak first-order phase
transitions in rotating neutron stars \citep{Erotquake1}.

For a  strong first-order phase transition ($\lambda<\lambda_{\rm crit}$) in neutron star core,
$\delta\overline{P}\longrightarrow 0$ implies
$r_{_{\rm S}}\longrightarrow r_{\rm _{S},min}$,
where $r_{\rm _{S},min}$ is a sizable fraction of the stellar
radius. The corequake, accompanying the phase transition, is a
large-scale phenomenon, with energy release $E_{\rm rel}\sim
10^{51}-10^{52}~$erg.
Existing numerical calculations were restricted to spherical
non-rotating neutron stars \citep{Migdal1979,HProsz1980,HaenProsz82,
Kaempfer1982,
Berezin1982,Berezin1983,Diaz1983,Berezhiani2003}.

In the present paper we calculate the energy release due to a
{\it strong first-order phase transition} in a  {\it rotating}
neutron star. The calculations are done using very precise 2-D
codes and a set of EOSs with strong first-order phase
transitions. We show, that similarly as in the case of a weak
first-order phase transition, studied in \citet{Erotquake1},
the energy released during a corequake depends only on the
excess of the central pressure of the metastable configuration
over $P_0$, and is to a very good approximation independent of
the angular momentum of collapsing star. Moreover, we show
that this property holds also for corequakes with initial
$P_{\rm c}<P_0$,  which require additional energy needed to
jump over the energy barrier.

The paper is organized in the following way. In Sections\ \ref{sect:EOS.theory}
and \ref{sect:families}
 we introduce notations and we describe general properties of the first-order
phase transitions in stellar core with particular emphasis on
the metastability and instability of neutron star cores.
Analytic models of the EOSs with first-order phase
transitions, allowing for very precise 2-D calculations, are
considered in Sect.\ \ref{sect:E.poly}, where we derive
generic properties of the energy release due to a first order
phase transition at the center of a rotating star. In Sect.\
\ref{sect:E.SLy} we present our results obtained  for a
realistic EOS of  normal phase, and we confirm remarkable
properties of the energy-overpressure relation (i.e.,
the independence from $J$), obtained in the previous section. 
In Sect.\ \ref{sect:decomposition} we discuss the decomposition of the energy release
into kinetic and internal energies. Sect.\ \ref{sect:Jumping} is devoted to the problem
of transition of one phase star into stable, two phase configuration through the energy barrier.
 Finally,  Sect.\ \ref{sect:conclusions} contains discussion of our results
 and conclusion.
%
\section{EOS with a strong first-order phase transition }
\label{sect:EOS.theory}
%
 At the densities under consideration, all constituents
of the matter are strongly degenerate, and temperature dependence
of the pressure and energy density can be neglected.
 At a given baryon density,  $n_{\rm b}$,  energy density of
 the N-phase of matter (including rest energies of particles which
 are matter constituents) is ${\cal
E}_{\rm _N}(n_{\rm b})$ and pressure $P_{\rm _N}(n_{\rm b})$.
The baryon chemical potential $=$ enthalpy per baryon  in the
N phase, is $\mu_{\rm _N}=(P_{\rm _N}+{\cal E}_{\rm _N})/n_{\rm
b}$. Similarly, one can calculate thermodynamic quantities for
the S-phase.

\begin{figure}[h]
\centering
\resizebox{3.0in}{!}{\includegraphics[clip]{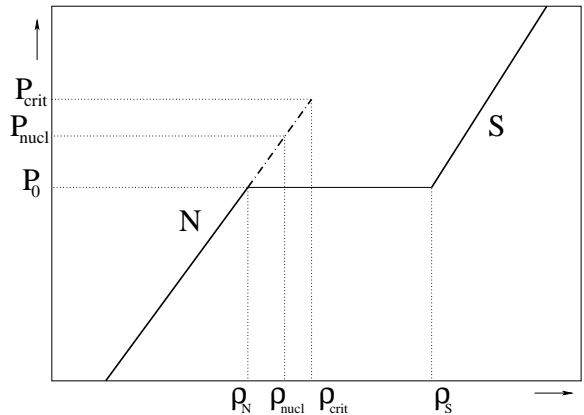}} \vskip
2mm\caption{A schematic representation, in the $\rho-P$ plane,
of an EOS with a strong first order phase transition,
$\rho_{_{\rm S}}/\rho_{_{\rm N}}>{3\over 2}(1+P_0/\rho_{_{\rm
N}}c^2)$. Solid segments: stable N and S phase (in
thermodynamic equilibrium). Dash-dot segment: metastable N
phase. The S phase nucleates at $P_{\rm nucl}$, which
  depends on the temperature and the compression rate. At $P_{\rm crit}$
nucleation of the S phase is instantaneous, because the energy
barrier,  separating the N phase from the S phase, vanishes. }
\label{fig:EOS1st}
\end{figure}
The value of $P_0$ is obtained from the crossing condition
$\mu_{\rm _N}(P)=\mu_{\rm _S}(P)$, which yields  also the values
of the matter densities, $\rho_{\rm _N}$ and $\rho_{\rm _S}$,
and the corresponding baryon densities, $n_{\rm _N}$ and
$n_{\rm _S}$, at the N-S phase coexistence interface. These
parameters are obtained assuming thermodynamic equilibrium. A
schematic plot of  EOS of matter with a strong first order
phase transition N$\longrightarrow$S,  in the vicinity of the phase transition
point, is plotted  in Fig.\ \ref{fig:EOS1st}.
Solid segment of the N-phase 
corresponds to the stable N-phase states. For
pressures above $P_0$, the N phase becomes metastable with
respect to the conversion into the S phase. The S-phase can
appear through the nucleation  process -  a spontaneous
formation of the S-phase droplets. However, an energy barrier
resulting from the surface tension at the N-S interface delays
the nucleation  for a time identified with a  lifetime of the
metastable state $\tau_{\rm nucl}$. The value of $\tau_{\rm
nucl}$ decreases sharply with $P>P_0$, and drops to zero at
some $P_{\rm crit}$, where the energy barrier separating the
S-state from the N-one vanishes. For $P>P_{\rm crit}$ the N
phase is simply unstable and converts with no delay into the S
phase. Calculations of $\tau_{\rm nucl}$ in dense neutron star core
metastable with respect to the pion condensation was performed
by \citet{HS82} and \citet{MT1992}. The case of nucleation of
 quark matter was studied by \citet{Iida1997,Iida1998}, while
 nucleation of the kaon condensate was considered by \citet{Norsen2002}.
 Discussion of $\tau_{\rm nucl}$
in an accreting or spinning down neutron star with increasing
$P_{\rm c}$ was  presented by \citet{Erotquake1}; a specific case of a
hot accreting neutron star was considered by
\citet{Berezhiani2003}.
\section{Families of neutron stars}
\label{sect:families}
For a given EOS, nonrotating hydrostatic equilibrium
configurations of neutron stars form a one-parameter family,
the parameter being, e.g., central pressure $P_{\rm c}$.
Notice, that $P_{\rm c}$ is preferred over $\rho_{\rm c}$,
because pressure is strictly continuous and monotonous in
stellar interior, while density can suffer discontinuities.
For an EOS with a first order phase
transition, and without allowing for metastability of the N
phase, neutron stars form two families: a family of
stars composed solely of the N-phase
 $\lbrace {\cal C}\rbrace$,
 and a family of those having an S-phase core
 $\lbrace {\cal C}^{\star}\rbrace$.

\begin{figure}[h]
\centering\resizebox{3.25in}{!}{\includegraphics[clip]{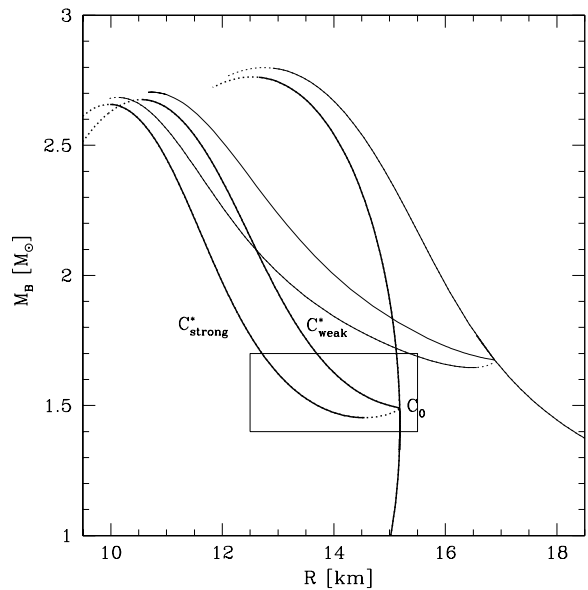}}
\caption{Baryon mass vs. equatorial radius for hydrostatic equilibrium
configurations calculated for three types
of EOS of dense matter, described in the text. Solid line - stable;
dotted  line -  unstable configurations. Thick lines correspond to the non-rotating
models, thin lines to the rigidly rotating
configurations with a fixed total stellar angular momentum
$J=1.2\times
GM^2_\odot/c$. }\label{fig:MbR_pol}
\end{figure}
It is known since longtime that configuration
$\lbrace {\cal C}^{\star}\rbrace$ with a small S-phase core
are unstable for $\lambda>\lambda_{\rm crit}\equiv {3\over
2}(1+P_0/\rho_{_{\rm N}}c^2)$ \citep{Seidov1971,Kaempfer1981,HZS,ZHS}.
Therefore, the topology of the complete
set of stable hydrostatic configurations of non-rotating
neutron stars, parametrized by $P_{\rm c}$, and plotted, e.g.,
in the mass-radius plane,   depends on the
value of $\lambda$.
Namely, for $\lambda<\lambda_{\rm crit}$, the sum of $\lbrace {\cal
C}\rbrace$ and $\lbrace {\cal C}^\star\rbrace$ is continuous,
while for $\lambda>\lambda_{\rm crit}$ it is not (i.e.,
the families $\lbrace {\cal C}\rbrace$ and $\lbrace {\cal C}^\star\rbrace$
are disjoint). Recently, it has been shown, that this
property is generic for an EOS with a phase transition,
and holds also for rigidly rotating neutron stars \citep{BBgen2006}.

An important global parameter for a hydrostatic equilibrium
configurations is its baryon mass, $M_{\rm b}$. It is defined as the baryon
number of the star, $A_{\rm b}$, multiplied by a 
``mass of a baryon'',
 $m_0$,  defined as the 1/56 of the mass of the
$^{56}{\rm Fe}$ atom: $m_0=1.6586\times 10^{-24}~$g.
 During evolution of an isolated neutron
star, including the phase transitions in its interior, $M_{\rm
b}$ remains strictly constant.

We define the equatorial radius of an axisymmetric neutron star as
the proper length of the equator divided by $2\pi$.
Examples of $M_{\rm b}$-$R_{\rm eq}$ curves for non-rotating
and rotating neutron stars without a phase transition
$\lbrace {\cal C}\rbrace$, with a weak first-order phase transition
$\lbrace {\cal C}^\star_{\rm weak}\rbrace$, and a strong
first-order phase transition
$\lbrace {\cal C}^\star_{\rm strong}\rbrace$, are shown in
Fig.\ \ref{fig:MbR_pol}. Dotted segments correspond to
unstable configurations (instability with respect to the
axisymmetric perturbations). As one sees,
 fast rotation significantly changes the $M_{\rm b}(R_{\rm eq})$
 dependence, e.g., by increasing the mass and the radius of the configuration
with $P_{\rm c}=P_0$ denoted by $C_0$.

General relations between models calculated for three
types of the EOS can be formulated. At a  given $M_{\rm b}$,
$R_{\rm eq}({\cal C}^\star_{\rm strong})
<R_{\rm eq}({\cal C}^\star_{\rm weak})<
R_{\rm eq}({\cal C})$. Moreover, maximum
allowable baryon masses satisfy
$M_{\rm b,max}({\cal C}^\star_{\rm strong})
<M_{\rm b,max}({\cal C}^\star_{\rm weak})<
M_{\rm b,max}({\cal C})$; the same inequalities are valid for
maximum allowable gravitational mass, $M_{\rm max}$.

\begin{figure}[h]
\centering\resizebox{3.25in}{!}{\includegraphics[clip]{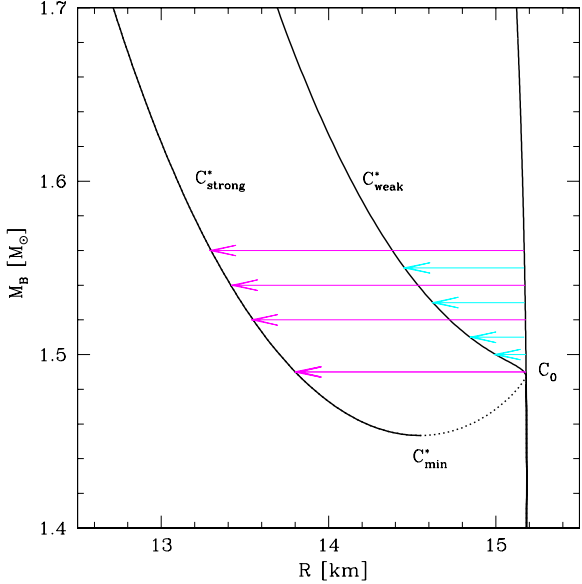}}
\caption{ Zoomed fragment of Fig.\ \ref{fig:MbR_pol}, in the vicinity
of the phase transition. For other explanations see the text.}
\label{fig:MbR_pol_zoom}
\end{figure}

Differences in the mass-radius behavior are most pronounced
in the vicinity of configuration ${\cal C}_0$, with $P_{\rm
c}=P_0$. This region, bounded by a rectangle
in Fig. \ref{fig:MbR_pol}, is shown in
Fig.\ \ref{fig:MbR_pol_zoom}, where the
 arrows connect configuration
with same $M_{\rm b}$. For simplicity, we first consider
non-rotating stars. For $\lambda<\lambda_{\rm crit}$,
configurations with $P_{\rm c}>P_0$ form a monotonous branch
${\cal C}^\star_{\rm weak}$. For $\lambda>\lambda_{\rm crit}$,
a segment with $P_0<P_{\rm c}<P^\star_{\rm {c,min}}$ consists
of configurations ${\cal C}^\star_{\rm strong}$ with ${\rm
d}M^\star/{\rm d}\rho^\star_{\rm c}<0$, which are therefore
unstable with respect to radial perturbations.
Therefore, for $\lambda>\lambda_{\rm crit}$ the stable branches
$\lbrace {\cal C}\rbrace$  and $\lbrace {\cal C}^\star\rbrace$
are disjoint.  In both cases
(``weak'' and ``strong''), central density jumps from
$\rho_{_{\rm N}}$ to $\rho_{_{\rm S}}$ when passing from
the ${\cal C}$ branch to the ${\cal C}^\star$ branch. All these properties
were derived long time ago for non-rotating neutron stars. Recently it has
been shown that they hold also for phase transitions in rigidly rotating
neutrons stars when the stationary axially symmetric
families $C{\cal }$ and ${\cal C}^\star$
contain configuration of  a fixed stellar angular momentum
$J$ \citep{BBgen2006}. The standard non-rotating case
corresponds to $J=0$.

\section{Calculation of the energy release}
\label{sect:E.num}
We restrict ourselves to axially symmetric, rigidly rotating neutron stars
in hydrostatic equilibrium. In what follows by ``radius'' we mean
the equatorial circumferential radius, $R_{\rm eq}$.

\begin{figure}[h]
\centering\resizebox{2.75in}{!}{\includegraphics[clip]{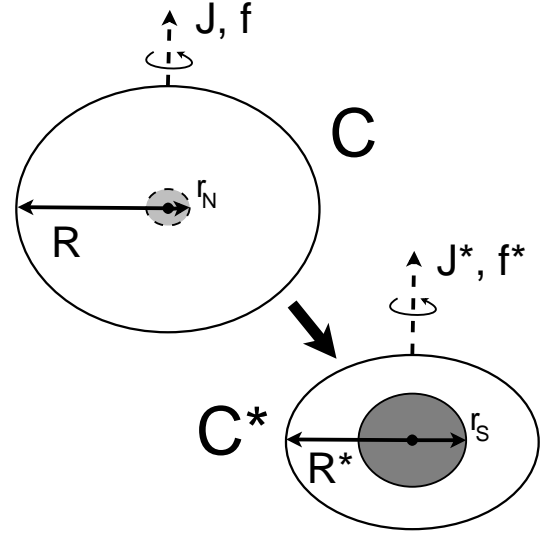}}
\caption{Transition from a one-phase configuration ${\cal C}$
 with a meta-stable core of radius $r_{\rm _N}$ to a
two-phase configuration ${\cal C}^\star$
 with  S-phase core with a radius $r_{\rm
_S}$. These two configurations have the same baryon mass $M_{\rm b}
=M_{\rm b}^\star$
and total angular momentum $J=J^\star$.}
\label{fig:CCstar}
\end{figure}

We assume that at a central pressure $P_{\rm c}=P_{\rm nucl}$
the nucleation of the S phase in an overcompressed core, of a
small  radius $r_{\rm _N}$ (of configuration ${\cal C}$)
initiates a strong first-order phase transition. This leads
to  formation of a large S-phase core of radius $r_{\rm _S}$
in a new configuration ${\cal C}^*$, as shown in Fig.\
\ref{fig:CCstar}.

We compare the hydrostatic equilibria
of neutron stars corresponding to the EOSs with and 
without the phase transition using the numerical library LORENE 
({\tt http://www.lorene.obspm.fr}), obtaining the axisymmetric, 
rigidly rotating solutions of Einstein equations as in \citet{Erotquake1}. 
The accuracy of the solution, measured with 
the general relativistic virial theorem \citep{GRV2} is typically $10^{-6}$.

The neutron-star models can be labeled by the
central pressure $P_{\rm c}$ (central density
is not continuous!) and rotational frequency $f=\Omega/2\pi$.
These parameters are natural from the point of view of numerical
calculations. But we can introduce  another
parametrization, more useful for other purposes. 
In order to study the stability
of rotating stars,  a  better choice is the central pressure,  $P_{\rm c}$,
 and the total angular momentum of the star,  $J$.

We additionally assume that the transition of the star from a  one-phase
configuration to the configuration with a dense core of
the S-phase takes place at fixed baryon
mass $M_{\rm b}$ (no matter ejection)
  and fixed total angular momentum of the star $J$
(loss of $J$ due to the electromagnetic or gravitational radiation is
 neglected).
The energy release during transition
${\cal C}(M_{\rm b},f)\longrightarrow {\cal
C}^\star(M_{\rm b},f^\star)$ is
therefore  calculated from the change of the stellar mass-energy
during this process,
\begin{equation}
E_{\rm rel}=c^2\;\left[M({\cal  C })-M({\cal C}^\star)\right]_{M_{\rm
b},J}~.
\label{eq:DeltaE}
\end{equation}

The precision of the determination of $E_{\rm rel}$ depends on the numerical accuracy
of calculations of the configurations with the same $J$ and $M_{\rm b}$. Taking into account
that $E_{\rm rel}$ is $3\div4$ orders of magnitude smaller than $M$ (for the overpressures
ranging from a few to 20  
percent) this precision is typically better than 1\%.
\begin{figure}[h]
\centering
\resizebox{3.25in}{!}{\includegraphics[clip]{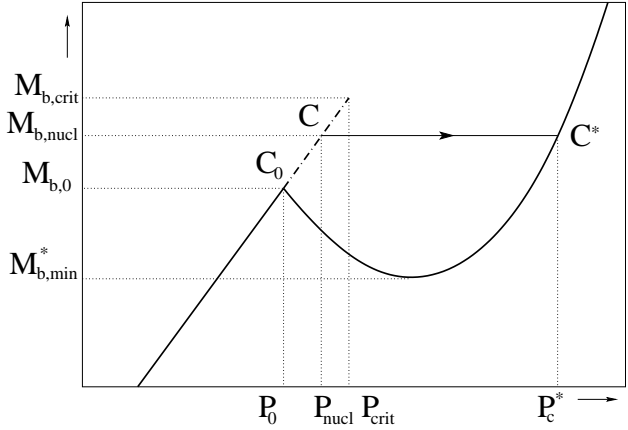}}
\caption{Total baryon mass $M_{\rm b}$ of hydrostatic stellar
configuration, versus central pressure  $P_{\rm c}$, at fixed
stellar angular momentum $J$, for the EOS depicted in Fig.\
\ref{fig:EOS1st}. Solid line denotes stable states,
dash-dot  line - the states which are
 metastable with respect to the
N$\longrightarrow$S transition. For a central pressure $P_{\rm
nucl}$ the S-phase nucleates in the super-compressed core of
configuration ${\cal C}$, and this results in a transition
${\cal C}\longrightarrow {\cal C}^*$ into a stable
configuration with a S-phase core and central pressure
$P^*_{\rm c}$. Both configurations  ${\cal C}$ and ${\cal
C}^*$ have the same baryon mass $M_{\rm b}$.}
 \label{fig:CCstarPc}
\end{figure}
\begin{figure}[h]
\centering \resizebox{\hsize}{!}{\includegraphics[angle=0]{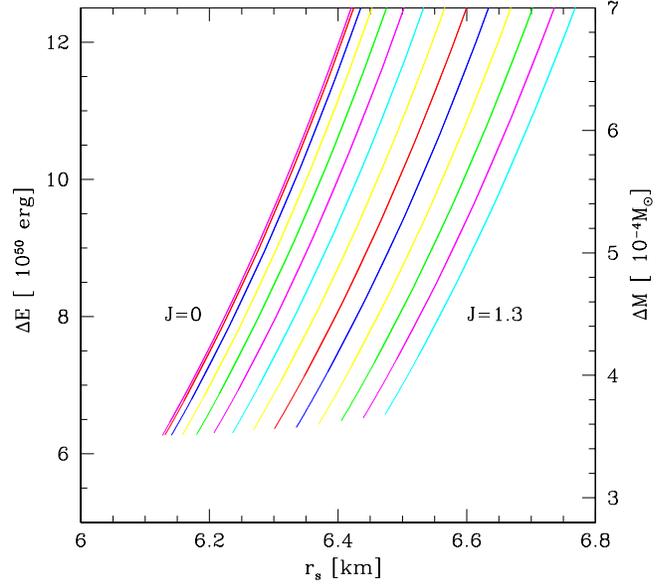}}
\caption{(Color online) The energy release due to the corequake  of
rotating neutron star as a function of the
equatorial radius of the S-phase core,
${r}_{\rm _S}$. Calculations are performed
for the EOS PolS1 from Table 1. Different  curves
 correspond to the different values of total angular momentum of
rotating star, fixed along each curve,
$J=\widetilde{J}\;GM^2_\odot/c=(0, 0.1,~\ldots,~1.3)\times
GM^2_\odot/c$, from the left-most  curve to the
right-most curve.
 }
\label{fig:EJ_rcore}
\end{figure}
\subsection{Energy release for polytropic EOSs}
\label{sect:E.poly}
\begin{table}[t]
\begin{center}
\caption {The parameters of the polytropic EOSs. The basic
EOS for the  N phase, named PolN,
 is $P=K_{\rm _N}c^2(n_{\rm b}/n_1)^{\gamma_{\rm _N}}$,
where $n_1=0.1~{\rm fm^{-3}}$ and $K_{\rm _N}=4.15\,10^{12}~{\rm
erg~cm}^{-3}$, and
$\gamma_{\rm _N}=2.5$.
Construction of the EOS for the S-phase  and of the first order
phase transition at $P=P_0$ is based on the Appendix A of
\cite{BBgen2006}. The transition point is the same for all three EOSs, and is defined by $P_0=3.686\,10^{34} {\rm erg~cm}^{-3}$,
$n_{\rm _N}=0.25\,{\rm fm}^{-3}$,
and $\rho_{\rm _N}=4.42\,10^{14}\,{\rm g~cm}^{-3}$. The density jump corresponding
to the phase transition is defined by the parameters
$\lambda=\rho_{\rm _S}/\rho_{\rm _N}$
and $\lambda_{\rm n}=n_{\rm _S}/n_{\rm _N}$, connected by the relation
$ \lambda=1+(\lambda_{\rm n}-1)(1+P_0/\rho_{\rm _N}c^2)$.
The S-phase EOS is a polytrope with the parameters $K_{\rm
_S}$ and $\gamma_{\rm _S}$. The non-rotating reference configuration
${\cal C}_0$  has $M=M_0=1.38\; {\rm M}_\odot$ and
$R=R_0=15.2\;{\rm km}$.
Phase transition in PolW1 is {\it weak} ($\lambda<\lambda_{\rm crit}$),
while those in PolS1 and PolS2 are {\it strong} ($\lambda>\lambda_{\rm crit}$). }
\begin{tabular}{c|c|c|c|c}
EOS &  $K_{\rm _S}$ & $\gamma_{\rm _S}$
& $\lambda$& $\lambda_{\rm n}$\\
 & $({\rm erg~cm}^{-3})$ & $({\rm erg~cm}^{-3})$&&\\
\hline
PolW &  $5.11\,10^{11}$ & 3.5  &$1.44$&$1.4$\\
\hline
PolS1 & $1.602\,10^{11}$ & 4 & $1.66$&$1.6$\\
\hline
PolS2  & $1.257\,10^{11}$ & 4 &$1.77$&$1.7$
\label{tab:poleos}
\end{tabular}
\end{center}
\end{table}
The
use of the polytropic EOSs for the N and S phases
not only guarantees very  high precision
of numerical calculation, but opens also a possibility
of the exploration of wide region of the parameter space.
Description of the polytropic EOSs
and their application to
relativistic stars with phase transitions was
presented in detail in our
previous publications in this series \citep{BHZ05,BBgen2006}.
\begin{figure}[h]
\centering
\resizebox{\hsize}{!}{\includegraphics[angle=0]{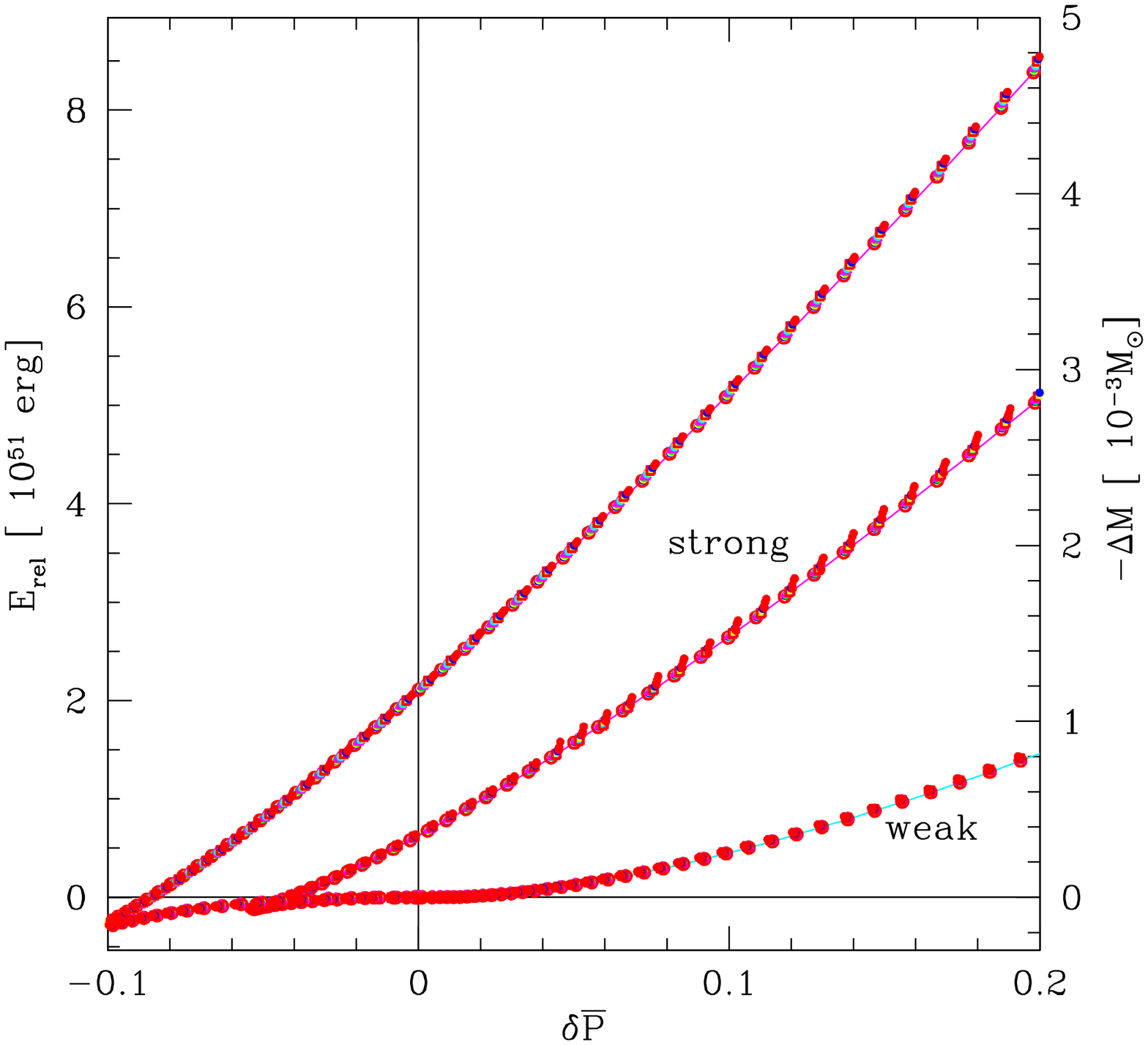}}
\caption{(Color online) The energy release due to the
mini-collapse of rotating neutron star as a function of the
overpressure $\delta\overline{P}$ of the N phase of the matter
in the center of the star,  for three  EOSs from Table 1. The
bottom curve is calculated for the PolW EOS wit a weak
first-order phase transition, so that $E_{\rm rel}(0)=0$. The middle curve
and the top curve are for the PolS1 and PolS2 EOSs,
respectively, with strong first-order phase transitions.
The points of different color correspond to the different values of total angular
momentum of rotating star,
 $J=\widetilde{J}\;GM^2_\odot/c=(0, 0.1,~\ldots,1.3)\times GM^2_\odot/c$.
For a given EOS,  results for all rotating configurations can be very well
approximated by a single curve, independent of $J$.
}
\label{fig:EJ_poly}
\end{figure}
In Fig.\ \ref{fig:EJ_rcore} we present the energy release
as a function of  ${r}_{\rm _S}$,
for several values of the angular
momentum of the metastable configurations ${\cal C}(J)$.
As we see the energy release
corresponding to a given value of $r_{\rm _S}$
depends rather strongly on the total angular momentum.
For example, the value of $E_{\rm rel}$ at $r_{\rm _S}=6.4\;$km,
for  $J=0$ (nonrotating star), is twice that for
$J=1.0\;GM^2_\odot/c$. Moreover,
let us notice  that the value of $E_{\rm rel}$ for minimum
$r_{\rm _S}$ (at fixed $J$) is (nearly) independent of $J$.
This reflects independence of $E_{\rm
rel}(\delta\overline{P}=0)$ from  $J$.

In Fig.\ \ref{fig:EJ_poly} we present the energy release
as a function of the overpressure of the metastable N phase in the center of the
metastable star ${\cal C}(M_{\rm b},J)$,
for several values of $J$.
As we already stressed, the value of $P_{\rm nucl}$ (or $\delta\overline{P}$)
 can be determined from microscopic considerations,
 combined with physical conditions prevailing
at the star center as well as
with the time evolution rate. Having determined
$P_{\rm nucl}$, we can  determine the energy release,
$E_{\rm rel}$,  due to the
corequake ${\cal C}(M_{\rm b},J)\longrightarrow {\cal C}^\star(M_{\rm b},J)$,
where the metastable one-phase configuration, and the final two-phase
configuration, have the same values of the baryon
mass  $M_{\rm b}$
and total angular momentum $J$.

As we see in Fig.\ \ref{fig:EJ_poly}, the energy release
in a  collapse of a rotating star is independent from the
angular momentum  of collapsing configuration, and  depends
exclusively on the degree of metastability of the N phase at the stellar
center (departure of matter from chemical equilibrium), measured by
the overpressure $\delta\overline{P}$. Consequently, to
 obtain the energy release associated with a corequake of a
 rotating neutron star, it is sufficient to know the value of
 $E_{\rm rel}$  for a non-rotating star of the same central
 overpressure.
\begin{figure}[h]
\centering
\resizebox{\hsize}{!}{\includegraphics[]{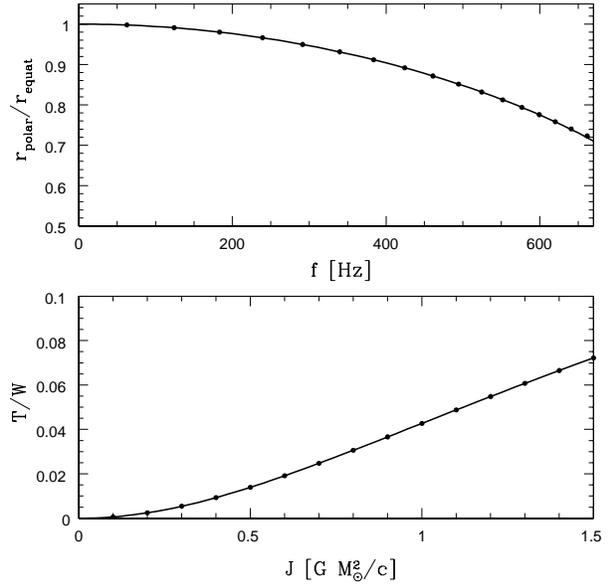}}
\caption{Top panel: the ratio of  polar radial coordinate to
the equatorial radial coordinate  ratio. Bottom panel:
the ratio of the  kinetic energy, $T$ and the absolute value
of the potential energy, $W$, for the reference stellar
configurations (central pressure $P_0$, Table 1)
consisting of the N phase of dense matter, described by the polytropic
EOS, PolN, of Table 1. Large dots correspond to the values of the total
stellar angular momentum, $J=\widetilde{J}GM^2_\odot/c=
(0.1,\ldots,1.5)\times GM^2_\odot/c$,
 which were used in Figs.\ \ref{fig:EJ_rcore},
 \ref{fig:EJ_poly}.
 }
\label{fig:C0.obl.TW}
\end{figure}
It should be stressed that the  configurations
${\cal C}(M_{\rm b},J)$ and $ {\cal C}^\star(M_{\rm b},J)$, considered in this
section, are the fast rotating ones, those with largest
$J$   are close to the  Keplerian (mass shedding) limit. This is
visualized in
 Fig.\ \ref{fig:C0.obl.TW}, where  we plotted the oblateness of the
 star and the kinetic to potential energy ratio.
  And still, in spite of
 fast rotation and large oblateness, the energy release
 is the same as in a non-rotating star of the same
 initial central overpressure.

 Summarizing,  a remarkable independence of $E_{\rm rel}$
 from $J$, obtained in \citet{BBgen2006}
 for $\lambda<\lambda_{\rm crit}$,
 when the small-core approximations were valid,  holds also
 for $\lambda>\lambda_{\rm crit}$, where perturbative
 arguments cannot be used.
%
\subsection{Realistic EOS - an example}
\label{sect:E.SLy}
%
In the present section we check the general validity
of results obtained for polytropic EOS and discussed
in Sect.\ \ref{sect:E.poly}, by
performing numerical calculations for realistic
EOSs with a strong first-order phase transition. For
the EOS of the crust we took
the model of \citet{sly}. The constituents of the N phase
of the core were neutrons, protons, electrons, and
muons. Nucleon component was described using the relativistic mean-field
model with scalar self-coupling, constructed by
\citet{ZM1990}. The values of the meson-nucleon coupling constants were
$g_\sigma/m_\sigma=3.122~$fm, $g_\omega/m_\omega=2.1954~$fm,
$g_\rho/m_\rho=2.1888~$fm. The dimensionless coefficients in the cubic and quartic
terms in scalar self-coupling were $b=-6.418\times 10^{-3}$  and
 $c=2.968\times 10^{-3}$, respectively. As an example of the S-phase
 we considered  the kaon-condensed
 matter. We constructed  a specific  dense matter model with
 a strong first-order  N$\longrightarrow$S
 transition implied by kaon condensation.
\begin{figure}[h]
\centering
\resizebox{\hsize}{!}{\includegraphics{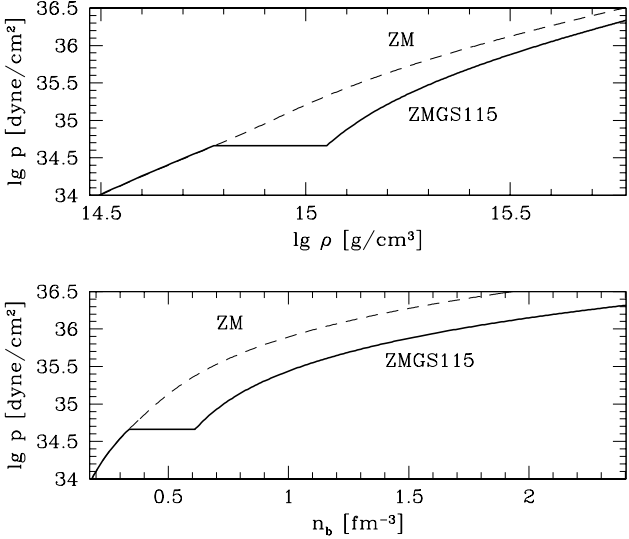}}
\caption{EOS with first-order phase transition,
due to kaon , described and used  in Sect.\ \ref{sect:E.SLy}.
in the present paper. The phase transition occurs
at $P_0=4.604\times 10^{34}~{\rm erg~cm^{-3}}$.
On the N-phase side, $n_{\rm _N}=0.3375~{\rm fm^{-3}}$ and
$\rho_{\rm _N}=5.973\times 10^{14}~{\rm g~cm^{-3}}$. On the kaon-condensed
S-side,  $n_{\rm _S}=0.6122~{\rm fm^{-3}}$ and
$\rho_{\rm _N}=1.125\times 10^{15}~{\rm g~cm^{-3}}$.
Therefore, $\lambda=\rho_{\rm _S}/\rho_{\rm _N}=1.884$,
which is greater than $\lambda_{\rm crit}=
{3\over 2}(1+P_0/\rho_{_{\rm N}}c^2)=1.629$.}
\label{fig:EOS_Kcond}
\end{figure}
Coupling of kaons to nucleons is described by the model of
\citet{GS-B1999}, with $U_K^{\rm lin}=-115~$MeV. Resulting EOS
 is shown in Fig.\ \ref{fig:EOS_Kcond}. The phase transition
 is the
{\it strong} one, with $\lambda>\lambda_{\rm crit}$ (see
the caption to Fig.\ \ref{fig:EOS_Kcond}).

In Fig.\ \ref{fig:EJ_Kcond} we show the energy release due to the
${\cal C}(M_{\rm b},J)\longrightarrow {\cal C}^\star(M_{\rm b},J)$
transition, versus overpressure. The
values obtained for different values of $J$ are marked with
different color and symbols. 
To an even better approximation
 than  for the polytropic models, all color points
lie along the same line. For  a given overpressure $\delta\overline{P}$,
the energy release  does not depend on $J$ of  collapsing metastable
configuration. This property holds for a broad range of
of stellar angular momentum, $J=(0.1,\ldots,1.0)\times G M^2_\odot/c$.
\section{Decomposition of the energy release}
\label{sect:decomposition}
%
As was already said, we restrict ourselves to the
stationary, axisymmetric states of rotating neutron stars.
In the Newtonian theory  energy of a rigidly rotating axially symmetric
star (body) is easily decomposed into kinetic energy of
rotation, $T$, and internal energy, $U$,
which is the the total energy of the star measured in the
stellar (body) reference
system. Kinetic energy of rotation is $T={1\over 2}J\Omega=
{1\over 2}I\Omega^2$, where the moment of inertia
 $I=J/\Omega$. On the other hand, the total energy
measured  in ``the laboratory frame'', $E$, is related to
$U$ by $U=E-J\Omega$.  For a rigid
rotation $U=E-{1\over 2}J\Omega$.

In general relativity, all kinds of energies sum up to give
the stellar gravitational mass, $M=E/c^2$, which is the source of
the space-time curvature. Therefore, the decomposition of $Mc^2$
into $T$ and $U$ is ambiguous. Here, we will use a standard
Newtonian-like formula, $T={1\over 2}J\Omega$, where both $J$
and $\Omega$ are well defined quantities \citep{FIP1986}.

\begin{figure}[h]
\centering \resizebox{\hsize}{!}{\includegraphics[angle=0]{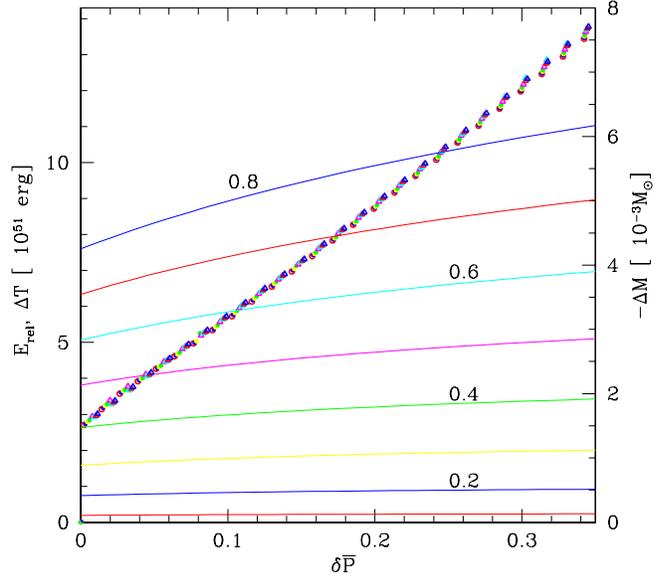}}
\caption{(Color online) The total energy release, $E_{\rm rel}$,
 due to the collapse of a
rotating neutron star, implied by kaon condensation,  as a function of the
metastability (overpressure) of the normal phase of the matter
in the center of the star for the EOS  in Fig.\ \ref{fig:EOS_Kcond}.
The points of different color correspond to the different values of
total angular momentum of rotating star. The results for all
rotating configurations can be very well
approximated by one curve.
Solid lines: kinetic energy increase, $\Delta T=T^\star - T$, for
fixed values of
$J=\widetilde{J}GM^2_\odot/c=
(0.1,\ldots,0.8)\times GM^2_\odot/c$. Bottom line is for
$\widetilde{J}=0.1$, and top line for $\widetilde{J}=0.8$.
 Note the different definition of the sign of $E_{\rm rel}=(M-M^\star)c^2$
and $\Delta T=T^\star - T$.
}
\label{fig:EJ_Kcond}
\end{figure}

The transition ${\cal C}(M_{\rm b},J)\longrightarrow {\cal C}^\star(M_{\rm
b},J)$ is accompanied by a spin-up of neutron star, and an
increase of its kinetic energy. Our results show, that while $E_{\rm rel}$
is to a very good approximation independent
from  $J$, the kinetic energy increase
$\Delta T\equiv
(T^\star - T)_{M_{\rm b},J}={1\over 2}J(\Omega^\star
- \Omega)_{M_{\rm b},J}$ grows rapidly with $J$. This is
visualized in Fig.\ \ref{fig:EJ_Kcond}.
%
The contribution to the energy release resulting from the decrease of the
internal energy of the star will be denoted by $E_{\rm rel}^{\rm (int)}$. It
is given by
\begin{equation}
E_{\rm rel}^{\rm (int)}(\delta\overline{P},J)
={1\over 2}J(\Omega^\star-\Omega)_{M_{\rm
b},J}+E_{\rm rel}(\delta\overline{P})~.
\label{eq:U-E-T}
\end{equation}
Therefore, the energy release in the star reference
system is greater than the total energy release, measured by a
distant observer. The difference increases rapidly with $J$,
see Fig.\ \ref{fig:EJ_Kcond}.
\section{Phase transition with climbing over the energy barrier}
\label{sect:Jumping}
\begin{figure}[h]
\centering\resizebox{3.25in}{!}
{\includegraphics[clip]{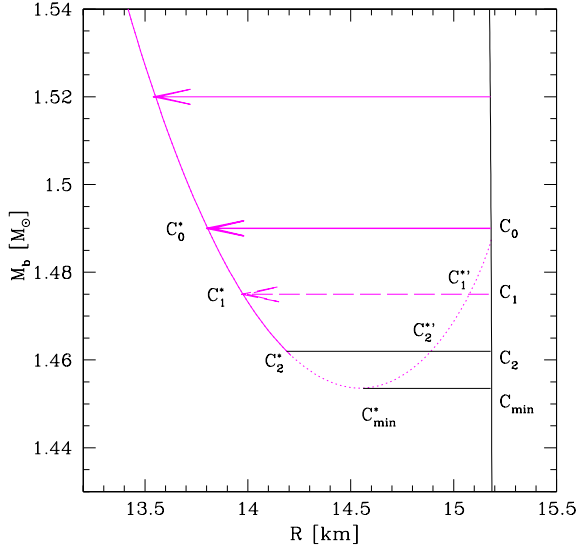}}
\caption{Three families of neutron stars in the $M_{\rm b}-R_{\rm eq}$
plane and the transition
 ${\cal C}_1\longrightarrow {\cal C}_1^\star$
 via jumping over the energy barrier.}
\label{fig:MbR_Jump}
\end{figure}

%
In the case of a strong first-order phase transition, configurations with
$P_{\rm crit}>P_{\rm c}>P_0$ are not the only ones which are metastable
with respect to the N$\longrightarrow$S transition.
In order to look for additional metastable N-phase configurations
we plotted, in Fig.\ \ref{fig:MbR_Jump},
 the vicinity of the reference configuration ${\cal C}_0$ in the
$M_{\rm b}-R$ plane: for $M_{\rm b}({\cal C}_0)>M_{\rm b}
>M_{\rm b}({\cal C}^\star_{\rm min})$ we have three
 equilibrium configurations with a given $M_{\rm b}$. 
Consider a triplet of equilibrium configurations
 ${\cal C}_1 {\cal C}_1^{\star\prime}{\cal C}^\star_1$.
 Two of them,  ${\cal C}_1$ and $ {\cal C}^\star_1$,
 are  stable (local minimum of
 energy), and one  ${\cal C}_1^{\star\prime}$ is unstable
 (local maximum of energy). Notice that the N-phase
 configurations on the ${\cal C}_0{\cal C}_{\rm min}$
 segment are characterized by {\it negative} overpressure
 (which might then be called ``underpressure'')
 $\delta\overline{P} = P_{\rm nucl}/P_0-1<0$.
\begin{figure}[h]
\centering\resizebox{3.25in}{!}
{\includegraphics[clip]{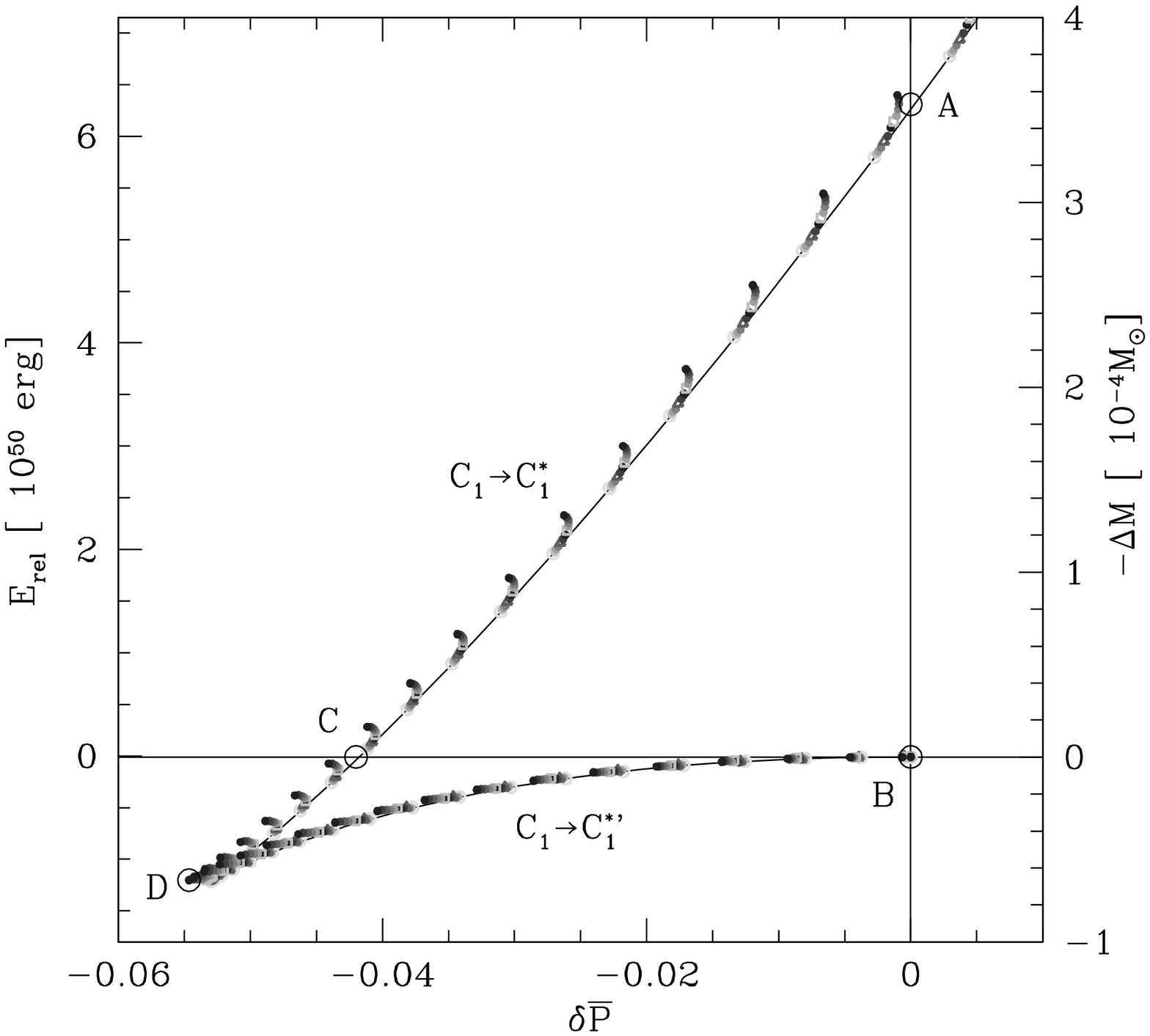}}
\caption{(Color online) Energy release, Eq.\ (\ref{eq:DeltaE}), versus
overpressure $\delta\overline{P}=(P_{\rm c}-P_0)/P_0$, for
transitions ${\cal C}_1\longrightarrow {\cal C}_1^\star$ and
${\cal C}_1\longrightarrow {\cal C}_1^{\star\prime}$
visualized in the $M_{\rm b}-R$ plane in Fig.\
\ref{fig:MbR_Jump}. Points of different color   correspond
to transitions with different angular momentum
$J=\widetilde{J}GM_\odot^2/c^2=(0, 0.1,
\dots, 0.9)\times GM_\odot^2/c^2$.  Solid line - results for $J=0$
(non-rotating stars). Point A corresponds to
${\cal C}_0\longrightarrow {\cal C}_0^\star$. Point B
corresponds to ${\cal C}_0\longrightarrow {\cal C}^{\star\prime}_0=
{\cal C}_0$, point D to ${\cal C}_{\rm min}\longrightarrow {\cal C}_{\rm
min}^\star$, and point C  to ${\cal C}_2\longrightarrow {\cal
C}_2^\star$.}
\label{fig:EJ_dP_minus}
\end{figure}

In Fig.\ \ref{fig:MbR_Jump} we indicated, by horizontal lines,
several examples of transitions between equilibrium
configurations with the same $M_{\rm b}$. To understand the
nature of these transitions, we plotted in Fig.\
\ref{fig:EJ_dP_minus} the energy release,
Eq.\ (\ref{eq:DeltaE}), associated with  a
transition between a pair of configuration, versus the {\it
negative} overpressure.
We see that the functional dependence $E_{\rm
rel}(\delta\overline{P})$, together with (quite precise)
independence from $J$, continue smoothly into the region of
negative $\delta\overline{P}$. Notice that point A
in Fig.\ \ref{fig:EJ_dP_minus} corresponds
to ${\cal C}_0\longrightarrow {\cal C}^{\star}_0$.  As far as the transitions
 ${\cal C}_1\longrightarrow {\cal C}_1^{\star\prime}$
 (segment BD in Fig.\ \ref{fig:EJ_dP_minus}) are concerned, they
  are  always associated with $E_{\rm rel}<0$, i.e. to make them the
 star should gain (absorb) energy instead of releasing it. The
 necessary energy input is ${\cal B}=|E_{\rm rel}|$, and it
 reaches a maximum at point D, corresponding to ${\cal C}_{\rm min}
 \longrightarrow {\cal C}^{\star}_{\rm min}$.
 Therefore, in
 order to get to ${\cal C}_1^{\star\prime}$
by forming a small S-phase core, the 
system has to climb over the energy barrier.
Then, configuration ${\cal
C}_1^{\star\prime}$ (which is unstable) can
collapse into ${\cal C}_1^{\star}$
with a large core, and this collapse is associated with an
energy release. In this way the star reaches the global minimum
of $M$ at fixed $M_{\rm b}$. However, this is the case
 provided the transition takes place above the
horizontal line ${\cal C}_2{\cal C}_2^{\star}$.

Summarizing, N-phase configurations on the ${\cal C}_0{\cal
C}_2$ segment are metastable with respect to transition to
large S-phase core configuration on the ${\cal C}^\star_0{\cal
C}_2^{\star}$ segment. However, such a transition requires
climbing over the energy barrier of a height ${\cal B}$. Let
consider a specific example with $\delta\overline{P}=-0.02$.
Using  Fig.\ \ref{fig:EJ_dP_minus}, we see that ${\cal B}\sim
10^{49}~$erg while $E_{\rm rel}=3\times 10^{50}~$erg.
Generally, for this model of phase transition we have $E_{\rm
rel}\gg {\cal B}$ for underpressures
$\delta\overline{P}>-0.03$. The excitation energy, $E_{\rm
exc}$, contained in radial pulsations, scales  as the square
of relative amplitude $\delta R/R$. For the fundamental mode,
$E_{\rm exc}\approx 10^{53}(\delta R/R)^2~$erg.
Therefore, $E_{\rm exc}$  exceeds  $10^{49}~$erg as soon as
$\delta R/R>0.01$, a condition that is easy to satisfy by an newly
born neutron star.

 However, a second obstacle for the collapse of configuration
 with a negative $\delta\overline{P}$ should be pointed out.
 Namely, apart from
the energy condition allowing climbing over an energy barrier, there is a
timescale condition: there should be enough time to form the
S-phase core. This latter condition may be more difficult
to fulfill than the former one, particularly if there is a
need to create strangeness, like in the kaon condensation or
in the formation of three-flavor u-d-s quark matter from a
deconfined two-flavor (u-d) state. Once again, favorable conditions for
${\cal C}_1\longrightarrow{\cal C}^{\star}_1$ with climbing
 over ${\cal C}^{\star\prime}_1$ could exist in a newborn neutron star.
A neutron star born in gravitational
collapse not only pulsates, with pulsational energy much greater than
${\cal B}$, but additionally a high temperature $\sim
10^{11}\;$K in the stellar core can allow for a  rapid nucleation
of the S-phase.
\section{Discussion and conclusions}
\label{sect:conclusions}
%
The most important result of the present paper is that the
total energy release, associated with a {\it strong} first-order phase
transition at the center of a rotating neutron star,  does
depend only on the overpressure at the center of the
metastable configuration  and is {\it independent from the star
rotation rate}.  This result holds even for fast stellar rotation,
when the star shape deviates significantly from sphericity,
and for overpressures as high as (10-20)\%.
 This property is of great practical
importance. Namely, it implies that the calculation of the
energy release for a given overpressure, requiring very high precision
2-D calculations to guarantee $M_{\rm b}=M_{\rm b}^\star$,  can be
replaced  by a simple calculation of non-rotating spherical stars.
The independence of the energy release on the rotation should be 
treated as a result of numerical calculations and is subject to the 
numerical accuracy of the stellar parameters determination.
Strictly speaking our numerical results indicate that, if the energy release
depends on the rotation, this dependence
is extremely weak. Namely, maximum deviation from the 
nonrotating value for a given overpressure 
is of the order of $1\%$, which actually 
is the precision of our numerical calculations. 

We studied stability of configurations with central pressure
below that for the equilibrium phase transition. If the
initial state of neutron star is excited, e.g., is pulsating,
then the  formation of a large  dense phase core is possible, but
it requires climbing over the energy barrier associated with
formation of a small core. Excitation energy has to be larger
than the height of the energy barrier. Additionally, if a
phase transition is connected with a change  of strangeness
per baryon, then the temperature has to be high enough to
make strangeness production sufficiently rapid. Such
conditions
might be realized in the cores of newly born neutron stars.

Note that the energy release $E_{\rm rel}\sim 10^{51}-10^{52}\;$erg, is an
absolute upper bound on the energy which can released
 as a result of a phase transition at the star
center.  The energy $\Delta E$ can be shared between, e.g.,  stellar
pulsations, gravitational radiation, heating of stellar
interior, X-ray emission from neutron star surface, and even
a gamma-ray burst.
%
\acknowledgements{
This work was partially
supported by the Polish MNiI grant no. 1P03D.008.27,
MNiSW grant no. N203.006.32/0450 and by the LEA
Astrophysics Poland-France
(Astro-PF) programme. MB was
also partially supported by the Marie Curie Intra-european
Fellowship MEIF-CT-2005-023644 within the 6th European
Community Framework Programme.}
%
%


\begin{thebibliography}{}
\bibitem[Bejger et al.(2005)]{BHZ05}
Bejger, M., Haensel, P., Zdunik, J.L., 
2005, MNRAS, 359, 699
\bibitem[Berezhiani et al.(2003)]{Berezhiani2003}
Berezhiani, Z., Bombaci, I., Drago, A., Frontera, F., Lavagno,
A., 2003, ApJ, 568, 1250
%
\bibitem[Berezin et al.(1982)]{Berezin1982}
Berezin, Yu.A., Dmitrieva, O.E., Yanenko, N.N., 1982,
Pisma v Astron. Zh., 8, 86
%
\bibitem[Berezin et al.(1983)]{Berezin1983}
Berezin, Yu.A., Mukanova, B.G., Fedoruk, M.N., 1983,
Pisma v Astron. Zh., 9, 116
%
\bibitem[Bonazzola \& Gourgoulhon(1994)]{GRV2}
Bonazzola, S., Gourgoulhon, E., 1994, 
Class. Quantum Grav., 11, 1775
%
\bibitem[Diaz Alonso(1983)]{Diaz1983}
Diaz Alonso, J., 1983, A\&A, 125, 287
%
\bibitem[Douchin \& Haensel(2001)]{sly}
Douchin, F., Haensel, P., 2001, A\&A, 380, 151
%
\bibitem[Friedman et al.(1986)]{FIP1986}
Friedman, J.L., Ipser, J.R., Parker, L.,
1986, ApJ, 304, 115
%
\bibitem[Glendenning(2000)]{Glend.book}
Glendenning, N.K., 2000, Compact Stars. Nuclear Physics,
Particle Physics, and General Relativity, Springer, Berlin
%
\bibitem[Glendenning \& Schaffner-Bielich(1999)]{GS-B1999}
Glendenning, N.K., Schaffner-Bielich, J., 1999,
Phys. Rev. C, 60, 025803
%
\bibitem[Haensel \& Pr{\'o}szy{\'n}ski(1980)]{HProsz1980}
Haensel, P., Pr{\'o}szy{\'n}ski, M., 1980, Phys. Lett., 96 B, 233
%
\bibitem[Haensel \& Pr{\'o}szy{\'n}ski(1982)]{HaenProsz82}
Haensel, P., Pr{\'o}szy{\'n}ski, M., 1982, ApJ, 258, 306
%
\bibitem[Haensel \& Schaeffer(1982)]{HS82}
Haensel, P., Schaeffer, R., 1982, Nucl. Phys. A 381, 519
%
\bibitem[Haensel et al.(1986)]{HZS}
Haensel, P., Zdunik, J.L., Schaeffer, R., 1986, A\&A, 160, 251
%
\bibitem[Haensel et al.(1990)]{HDPopov1990}
Haensel, P., Denissov, A., Popov, S., 1990,
A\&A, 240, 78
%
\bibitem[Haensel et al.(2007)]{NSB1}
Haensel, P., Potekhin, A.Y., Yakovlev, D.G.,
2007, Neutron Stars 1. Equation of State and Structure, Springer, New York
%
\bibitem[Iida \& Sato(1997)]{Iida1997}
Iida, K., Sato, K., 1997,  Prog. Theor. Phys., 98, 277
%
\bibitem[Iida \& Sato(1998)]{Iida1998}
Iida, K., Sato, K., 1998, Phys. Rev. C., 58, 2538
%
\bibitem[Kaempfer(1981)]{Kaempfer1981}
Kaempfer, B., 1981, Phys. Lett. 101B, 366
%
\bibitem[Kaempfer(1982)]{Kaempfer1982}
Kaempfer, B., 1982, Astron. Nachr., 303, 231
%
\bibitem[Migdal et al.(1979)]{Migdal1979}
Migdal, A.B., Chernoutsan, A.I., Mishustin, I.N., 1979,
 Phys. Lett. 52B, 172
%
\bibitem[Muto \& Tatsumi(1990)]{MT1992}
Muto, T., Tatsumi, T., 1990, Prog. Theor. Phys. 83, 499
%
\bibitem[Norsen(2002)]{Norsen2002}
Norsen, T., 2002, Phys. Rev. C, 65, 045805
%
\bibitem[Seidov(1971)]{Seidov1971}
Seidov, Z.F., 1971, Sov. Astron.- Astron.Zh., 15, 347
%
\bibitem[Weber(1999)]{weber.book}
Weber, F., 1999, Pulsars as Astrophysical Laboratories for Nuclear
and Particle Physics, IoP Publishing, Bristol \& Philadelphia
%
\bibitem[Zdunik et al.(1987)]{ZHS}
Zdunik, J. L., Haensel, P., Schaeffer, R., 1987, A\&A, 172, 95
%
\bibitem[Zdunik et al.(2006)]{BBgen2006}
Zdunik, J. L., Bejger, M., Haensel, P., Gourgoulhon, E., 2006,
A\&A, 450, 747
%
\bibitem[Zdunik et al.(2007)]{Erotquake1}
Zdunik, J. L., Bejger, M., Haensel, P., Gourgoulhon, E., 2007,
A\&A, 465, 533
%
\bibitem[Zimanyi \& Moszkowski(1990)]{ZM1990}
Zimanyi, J., Moszkowski, S.A., 1990,  Phys. Rev. C, 42, 1416
\end{thebibliography}
\end{document}